# An Open Science Platform for the Next Generation of Data

Latanya Sweeney and Merce Crosas

Harvard University[1]

## Abstract

Imagine an online work environment where researchers have direct and immediate access to myriad data sources and tools and data management resources, useful throughout the research lifecycle. This is our vision for the next generation of the Dataverse Network: an Open Science Platform (OSP). For the first time, researchers would be able to seamlessly access and create primary and derived data from a variety of sources: prior research results, public data sets, harvested online data, physical instruments, private data collections, and even data from other standalone repositories. Researchers could recruit research participants and conduct research directly on the OSP, if desired, using readily available tools. Researchers could create private or shared workspaces to house data, access tools, and computation and could publish data directly on the platform or publish elsewhere with persistent, data citations on the OSP. This manuscript describes the details of an Open Science Platform and its construction. *Having an Open Science Platform will especially impact the rate of new scientific discoveries and make scientific findings more credible and accountable.*

**Keywords**: open science, open data, data reuse, data citation, big data, data science, provenance, data privacy, real-time data analysis, data sharing

---

## Vision

We envision a computational and data management research platform that gives researchers direct and immediate access to myriad data sources and tools, useful throughout the research lifecycle. Figure 1 offers a summarizing depiction. For the first time, researchers will be able to seamlessly access and create primary and derived data from a variety of sources: prior research results, public data sets, harvested online data, physical instruments, private data collections, and even data from other standalone repositories. Researchers can recruit research participants and conduct research directly on this _Open Science Platform (OSP)_, if desired, using readily available tools. Researchers can create private or shared workspaces to house data, access tools, and computation and can publish data directly on the platform or publish elsewhere with persistent, data citations on OSP. _The Open Science Platform will especially impact the rate of new scientific discoveries and make scientific findings more credible and accountable_.

Examples may include a neuroimaging researcher who _seeks public assistance_ to manually proofread and correct results from automated processing of nanometer resolution brain tissue images, generated at the _big data_ rate of ½ terabyte per day.  A public health researcher wants to integrate 10 years of _privacy-sensitive_ health data with environmental data and then _openly share_ derived data with other researchers.  An astronomer learns of a new _analytical tool_ used _in another discipline_ and wants to use it on an _existing standalone data store_, but the database does _not have a tools API_ or the new algorithm. An educational researcher who captures _privacy-sensitive_ student demographics, click streams and social networks from Massive On-line Open Courses (MOOCs) needs to find _the best big data storage schemes_ for analyzing these different kinds of data and wants to _explore big data in real-time_ to predict performance and, perhaps, to customize course materials on the fly. A medical researcher _seeks research participants_ for a clinical trial and wants to reduce the time to locate and enroll study participants from 18 months to one hour. A social scientist working with United National Global Pulse found twitter data to be an accurate indicator of people's stress about food, fuel, finance and housing and wants to _cite the sample of the data stream_ used and enable others to _replicate findings_.  A biologist _searched a data repository_ and found useful data from which interesting discoveries resulted, so he needs to _know the provenance of the data_ –its origins and how it has been curated or possibly altered.

The Open Science Platform needs to provide its resources openly to researchers across different disciplines. It must work naturally with researchers, not be rigid or dictatorial, but rather enable researchers to do the work they want to do whenever and wherever, throughout the research lifecycle, from idea exploration through dissemination of results.  OSP must be reliable and robust and financially sustainable. Researchers must be able to trust its tools, storage, privacy, and citation features.

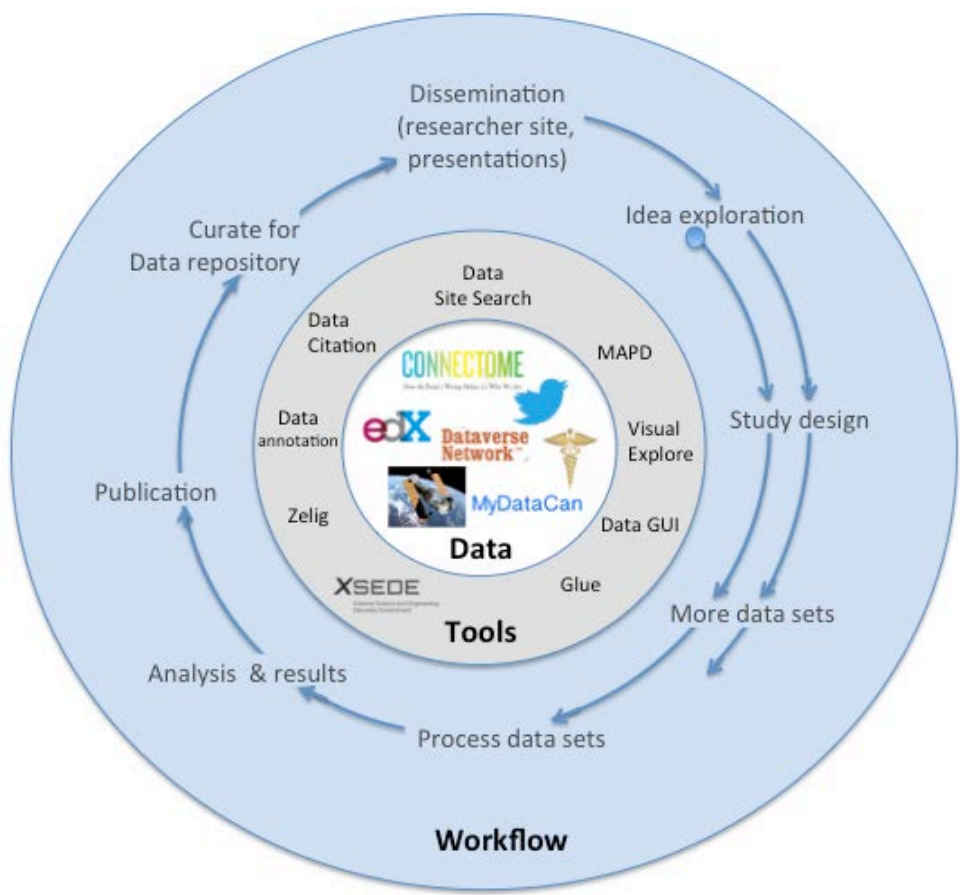

**Figure 1. Research Lifecycle**

## Need

The Open Science Platform will transform research by making more diverse kinds of data available for exploring ideas and replicating and rapidly building on prior work.

Society generates data on a scale previously unimagined.  There are 1.2 trillion yearly searches on Google, 1.1 billion smartphone subscribers worldwide, 400 million tweets per day, and 7 petabytes of photo content added to Facebook every month [1,2]. The United Nations Global Pulse Initiative describes efforts that used these kinds of data to derive results about unemployment and public health that were as accurate and acquired significantly faster than traditional compilations [3].

Scientific understanding is improving with data too. Our lives, organizations, societies, and universe are beginning to be transformed by advances in the likes of computational social sciences [4, 5, 6], biomedical science and genomics [7], astronomy [8], education [9], public health [10], and biomedical imaging [11].

Size, opportunity or sensitivity keeps these data inaccessible to other scientific disciplines and researchers. Limiting data to a few chosen researchers or driving data that enables science back into the vaults of private industries --- wireless carriers, credit card companies, hospitals, etc. --- would risk losing a multitude of opportunities to better understand our lives, and societies.

Data sharing maximizes the value of collected data and promises follow-up studies using existing data.  Sharing also minimizes duplicative data collection, which in turn reduces research costs and lowers the burden on human research participants. If science is to be progressive and self-correcting, then data, not just summary conclusions, must be open to independent scrutiny [12].

The Federal Government also wants to spur scientific breakthroughs and economic advances by making research results more available to innovators. Under a new White House directive, all federal agencies with R&D budgets in excess of $100M will have to develop their own public access policies that will "ensure the public can read, download, and analyze in digital form" published works arising from federally-funded research [66].

Future researchers using the Open Science Platform will be able to rapidly explore new ideas and formulate and test hypotheses quickly because OSP minimizes the overhead of data collection, acquisition and handling, even the handling of privacy-sensitive data. OSP will enhance and augment research capability, and help improve science by allowing easy replication of research results. The ultimate economic benefits are lower research costs and a stronger economy, because the time to transition research results into practice will be less. A researcher's personal potential increases as he has many more opportunities for new discoveries and making major scientific contributions.

## State of the Art

A huge quantity of digital research data exists solely as files on the computers of individual researchers, even in cases where the impact of the research based on the data has been dramatic. If the community does not take steps to preserve it, it will be lost forever. It needs to be identified, located, assessed, acquired, processed, preserved, and shared to extend its useful life and preserve it against computer failure [67].

Over the last 7 years, different web archives emerged for sharing data studies. Most are discipline-specific with the quality and nature of management differing  (e.g., geophysics [69], biosciences [71], earth sciences [72], clinical trials [68] and medicine [70]).

The leading, most comprehensive archives are capable of publishing, citing and sharing research data and they tend to be powered by the same software, The Dataverse Network, an open-source application founded at Harvard in 2006 that has been installed and used in a multitude of Universities around the world as a research data-sharing platform for those institutions (e.g., a consortium of Universities in the Netherlands and Fudan University in China). The Harvard instance of the Dataverse Network hosts more than 51,000 studies having 719,000 files, the world's largest collection of social science research data.[2]  The Astronomy Dataverse Network and all other disciplines at Harvard are merging into a main Harvard instance [6, 14].

While the current Dataverse and similar data archives make important contributions, they need a new platform, one able to accommodate the full spectrum of challenges researchers face, e.g., difficulties accessing, storing and citing big data, handling privacy-sensitive data, interoperating with other repositories and large computation facilities and becoming financially sustainable.

---

[2] http://theData.org

The next generation of the Dataverse will be a potent organizing mechanism, as a primary source and catalyst, for creating fundamental knowledge for open science platforms. The missing element is a basic understanding of how researchers work. Any OSP must address this core problem using researcher centered, holistic design that brings together researchers from a variety of disciplines with information technology engineers and computer scientists and those experienced in curating, citing and managing research data.

The key to success is to study real researchers engaged in real research in order to make sure the developed system responds to actual needs. Of course constructing an OSP is not an ethnographic study of research practice, but one of properly augmenting research functions with technology; it is not a technical model either, but one that integrates all aspects of function and participation in the research lifecycle.

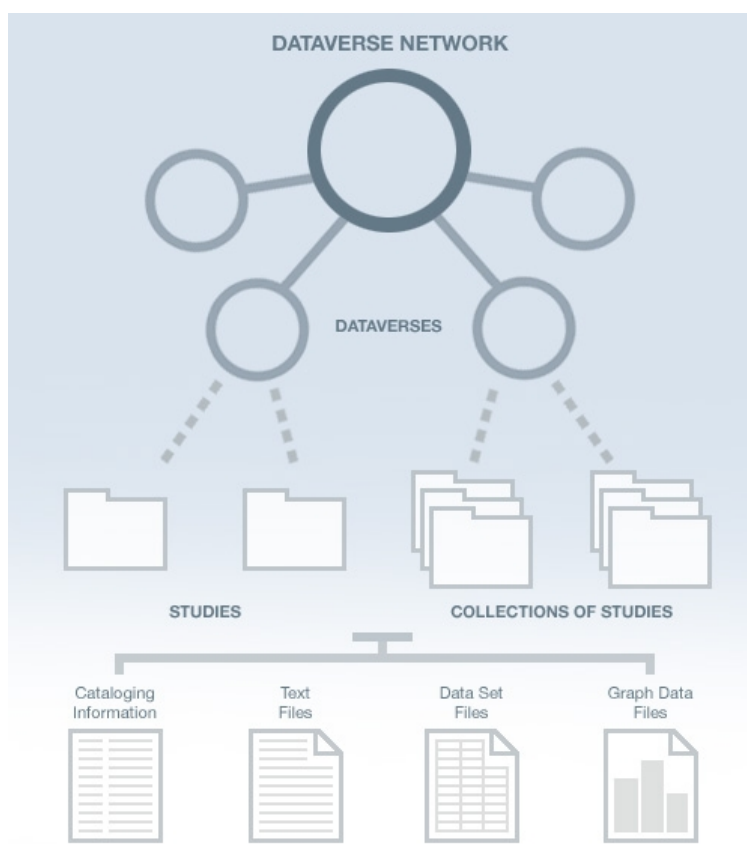

**Figure 2. Organization of a Dataverse Network**

As a result, the OSP serves as a platform for enabling and enhancing data science for all researchers and all data-driven research, leveraging a variety of skills, tools, and database technologies across disciplines.

All the features described in this paper about the construction of an OSP are generally applicable to a host of open science platforms. The ultimate goal is to make leap-forward breakthroughs, not incremental advances to a specific platform.  By taking an integrative and complex systems approach, we describe a broad cross-disciplinary approach to constructing an OSP that draws on design partners as subjects and as researchers.

## The Dataverse Network

One of the existing projects we build upon in the feature design of the Open Science Platform is the Dataverse Network, the leading software for sharing research data and allowing replication of previous work.

A Dataverse Network hosts multiple *dataverses*. Each dataverse is a "virtual" data archive customizable by its owner (or even embedded in the owner's personal web site), and it contains studies or collections of studies, and each study contains cataloging information that describes the data, in addition to the research data files and files (e.g. documentation and code).  Figure 2 has a graphical depiction.

The Dataverse Network provides: searchable descriptive metadata about the data and metadata extracted automatically from data files; user-friendly web interface to deposit, manage and preserve data; enough information (data, documentation and code) to reproduce and validate a scientific study without contacting the author; formal data citation, including a persistent global url and data fingerprint ,with attribution to data authors and distributors; and, various levels of access: fully open, data agreements, access with approval. [15,17,18]. A key feature is its incentive structure --researchers and data authors get credit, publishers and distributors get credit, and affiliated institutions get credit [15, 16].

In earlier work, Gary King and Merce Crosas identified eight core social/institutional responsibilities for a data-sharing infrastructure [15,16]:

*1. Recognition*. Citation credit should be apportioned both for the original article and separately for the data.

*2. Public Distribution*. Data distributed with an article should become accessible by the scientific community without having to obtain permission from the author for each use. Having to obtain permission from an author to read a published article, such as by agreeing ex-ante not to criticize it in print, is so obviously unacceptable it no longer occurs. The same should be true for data. Science requires the transmission of information through public means, not private understandings.

*3. Authorization*. Although free and open access to data is ordinarily preferable, it is neither always feasible nor necessary for the purpose of guaranteeing the public transmission of information. Those who wish to access the data can reasonably be asked to fulfill whatever authorization requirements the original author needed to meet in order to acquire, distribute, and archive it in the first place. This may include signing a licensing agreement (such as agreeing to respect confidentiality pledges made to research subjects), signing the equivalent of a guest book, being at an institution with a membership to the right archive or even paying a fee. And different requirements may apply to different portions of a data set.

*4. Validation*. Editors, their designees, and future researchers need to be able to check that a specific set of data exists, even if they have not met the authorization requirements.

*5. Verification*. Journals and researchers need to ensure that data associated with each published article is frozen and cannot be changed without detection, regardless of transformations –i.e., from SPSS to Stata to R, from a PC to a Mac to Linux, and from 8 inch magnetic tape to 5.25 inch floppies to a DVD.

*6. Persistence*. Researchers decades from now need to be able to find the data, access it, validate that it is the data set associated with the article in question, no matter what changes appear in methods of data distribution and network access, data storage formats, statistical and database software, operating systems, and computer hardware.

*7. Ease of Use*. Tasks must be simple and easy to use and follow.

*8. Legal Protection*. Publishers have well-honed procedures for dealing with copyright and liability issues for printed matter, but these standard copyright assignment forms do not cover acquiring and distributing data off the printed page.

The Open Science Platform incorporates these 8 responsibilities.  OSP also incorporates statistical packages, such as the R-like statistical package called Zelig, an online tool available in the Dataverse Network.  These packages will provide an API for others to build analytical tools for the OSP platform.

The overall feature development for managing research data in the Open Science Platform is to start with the data management features inherited from Dataverse, then expand to enable the same with privacy-sensitive datasets, then with big data, and finally, using high performance computing (as shown in Figure 3).

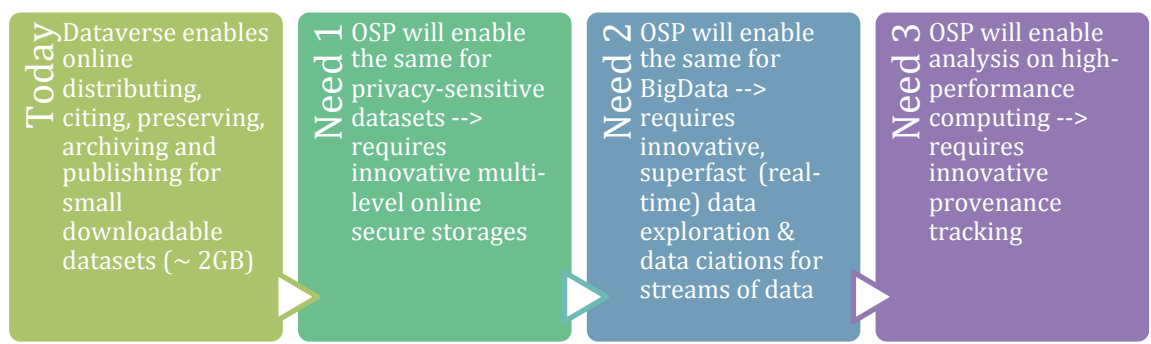

**Figure 3. Development of data management features in the Open Science Platform.**

⌘ Architecture of The Dataverse Network

Even though an OSP uses a different architecture, we describe the Dataverse Network's architecture for comparison. It has a multi-tier Java enterprise architecture, which leverages open-source technologies to:1) index the metadata to provide fast searchers (Lucene Index server), 2) register persistent identifier to a global registry (Handle server), 3) analyze quantitative data sets (Zelig and RServe), 4) inter-operate with other systems by exporting and harvesting metadata (OAI-PMH server and client), 5) copy the data to multiple locations (LOCKSS server), and 6) PostgreSQL for storing user data and research study medatata.

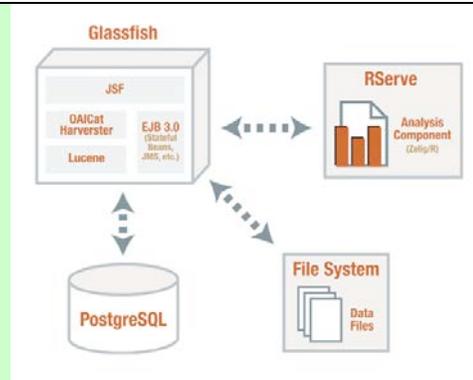

**Architecture of Dataverse Network.**

## Development

Developing the Open Software Platform involves creating the scientific and engineering knowledge base that enables systematic development of researcher-centered platforms to support researchers and enable research throughout the research lifecycle. We learn what's important by involving researchers who are engaged in their own research.

For example, Popper proposed his theory of scientific discovery that epistemologically defined the conventional view of today's research process [73]. Popper characterized the research process as starting with a clearly formulated research problem followed by different steps that sequentially lead to its solution. This is the formal outline adopted and accepted more or less unquestionably by almost all. Most likely Popper's definition is the result of observing documentation from the practice –academic publications – rather than creating it, because for most researchers, the research process is not linearly sequential, but is a cyclic process of exploring an idea, formulating a hypothesis, testing and analyzing it, and then repeating the process. On each cycle, the research improves based on *feedback* from the previous cycle.

So, an OSP must be designed for researchers' iterative approach and likewise, an OSP's design improves iteratively as well. Almost subliminally we view the development of an OSP as a model built and continuously refined from training, experience, and the implicit understanding of advancing through the steps of a research study of itself.

For the Open Science Platform to be successful, the engineer and the participating researcher must have an acute understanding of the needs of researchers. Indeed, in preparing this manuscript we did not simply rely on our collective years of research experience, but instead we reached out to researchers in other disciplines through a series of meetings to discuss issues they were experiencing as they conducted their research. The results are the research scenarios (shown in insets) that represent the need for and the environment of the Open Science Platform. They serve as motivators and illustrations throughout this manuscript.

| | Development Thrusts | | | | | | |
|---|---|---|---|---|---|---|---|
| | Core & APIs | Big Data | Data Explore | Privacy Sensitive | Provenance Tracking | Publishing & Citation | Sustain |

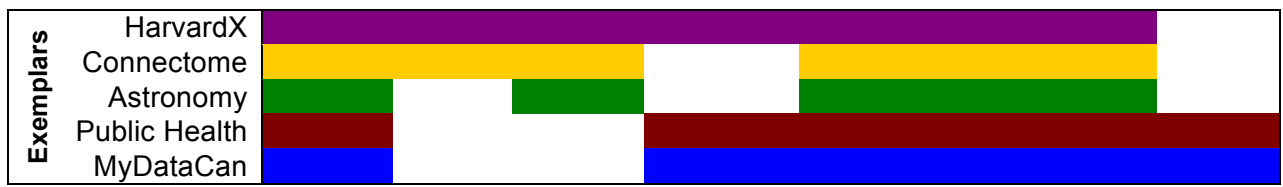

**Figure 4. Development thrusts intersecting with data research exemplars.**

Our approach centers on seven development thrusts: core platform and APIs; big data; data exploration; provenance tracking; publishing and citation; privacy-sensitive data; and, sustainability that integrates five crosscutting data partners, as exemplar use cases: Harvard X; connectome; astronomy; public health; and, MyDataCan. Figure 4 shows how the thrusts and exemplars intersect.

Thrusts involve engineering specific features of the OSP platform. They are:

1) <u>Core Platform & APIs</u> –The Open Science Platform must be able to perform basic data management operations, have extensible APIs for use with third-party tools and standalone data repositories, and adhere to the eight responsibilities inherited from the Dataverse Network (listed previously).

2) <u>Big Data Storage Systems</u> –The Open Science Platform must provide researchers the ability to efficiently store, analyze and explore big data –data and data streams that are far too large to copy locally and work with using traditional tools.

3) <u>Data Exploration Tools</u> –The Open Science Platform must have tools that work naturally with researchers in exploring data and conducting analyses.

4) <u>Privacy-Sensitive Data Handling</u> –The Open Science Platform must support the storage, analysis and handling of sensitive personal information.

5) <u>Provenance Tracking</u> –The Open Science Platform must provide researchers with a record of lineage and authenticity of data.

6) <u>Publishing and Citation</u> –The Open Science Platform must provide a means of citing data, even samples drawn from data streams, and maintaining meta-data so that research results can be replicated.

7) <u>Financial sustainability</u> –The Open Science Platform must operate with financial independence from sponsors in the long-term.

There is a naturally occurring integration across the technological dimensions of the thrusts. A researcher exploring big stores of sensitive data suggests an integrated architecture (e.g., the HarvardX inset describes), for example.

**Data Partner: HarvardX**

<u>HarvardX</u> is the name given to the online and hybrid online and on-campus courses offered by Harvard University using the edX technology. While most of the attention to the technology has centered on its ability to cross lines of space and time for education, the technology is also designed to provide unprecedented amounts of data for researchers.

The data comes in a number of different forms. Some of it is fairly static demographic or survey data, which differs from the data traditionally gathered in educational research only by the number of students involved, which may range from thousands to hundreds of thousands, depending on the class. More interesting is the interaction data, gathered at both the course and the server level, which records information about the interactions of each student with any material in the course. Thus, the HarvardX team led by Professor Waldo has gathered streams of data on how students watch lectures, engage with textbooks, try assignments, or do pop-quizzes, including how long they spend on each activity and how often they retry one activity or another. Finally, the HarvardX team has access to all of the data generated by the discussion forums that are integral to the courses. A typical course has 50,000 students.

These data are generated at a high rate, and can be used to predict performance and, perhaps, to customize course materials on the fly. A problem is presenting these data to education researchers and teachers, both in how they can best see the results in a way that can allow them to make sense of the scale and how to customize the views of streams of data to allow new insights. Also, this data is highly personalized and protected by FERPA and other privacy regulations, posing restrictions on sharing with researchers.

In developing an Open Science Platform, Harvard X researchers allow us to test different storage schemes with different kinds of data to find the best schemes for desired analyses. Experiments include analyzing different graph engines; assessing privacy risks; and, assessing data anonymizations.

## 1. Core Platform & APIs

This development thrust focuses on the basic operations of the OSP and on extensions to third parties. It is far less innovative than the other thrusts. The Open Science Platform goals for this thrust are to support the ability for researchers to work in unstructured ways using open access to data and tools on the OSP.

The Open Science Platform will include: extensible data storage from fixed-length files to streaming big data, connectors to distributed storages of data of big and fixed size data in local and external clouds, and extensible data exploration and analysis tools (e.g., for crowd-sourcing annotations and multi-dimension visual exploration in real-time). Figure 5 provides an architectural overview.

The OSP architecture has three tiers.  The Public Tier (shown on top) offers public facing tools and applications for researchers to interact with the platform. The named boxes across the top include several tools that this thrust will provide, namely: _annotation_, _visual exploration_, _statistical and privacy tools_, a _dashboard web interface_ that serves as a central online entrance point for researchers, a set of applications and web interfaces for people to use with _private repositories_ of their own personal information, and a _batch facility_ for uploading lots of data or jobs as a unit.

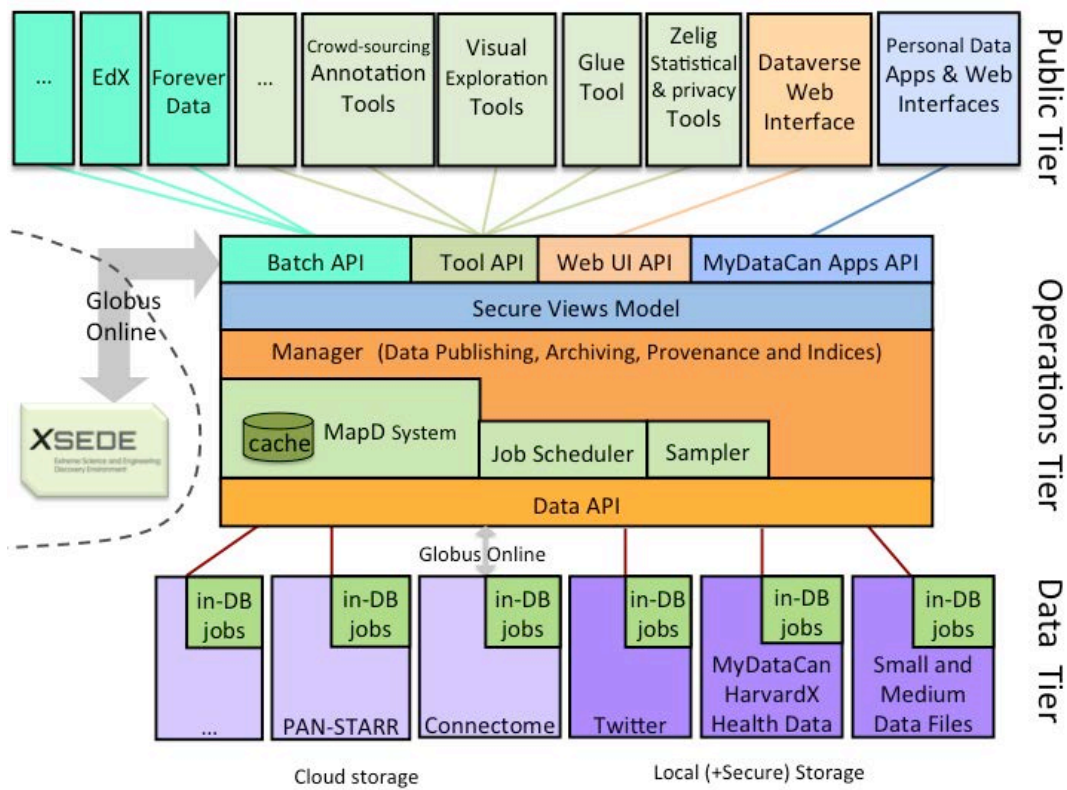

**Figure 5. Architecture of the Open Science Platform.**

⌘ Related Activities Underway

The application components of the Dataverse Network (similar to the OSP's Manager) include support for: Publishing model (draft –review-release workflow for contributors, curators and data owners); Archival functions (no deletion of studies after published, only deaccession); Versioning model; Data citation generation, with persistent identifier registration (Handle Server); Metadata indexing/searching (Lucene server); Ingest model (metadata/variable extraction from data files, UNF calculation); XML metadata generation (DDI, Dublin Core, MARC); OAI Harvester client/server; LOCKSS server (data duplication in multiple locations); Authentication & Authorization service: user & groups, user roles, object permissions, guestbook & terms of use, single sign-on (Shibboleth); RESTful API for: search, get metadata, download data; and, SWORD API.

The Dataverse team is already building an abstraction layer to Harvard's research-computing cluster that will page out to Massachusetts Green High Performance Computing Center, and any other cloud, and have almost completed work to connect Dataverse to Amazon's cloud.

The rules that govern interactions (or "APIs") between the Public and Operations Tiers will be openly available so that anyone can develop software for use on the OSP. Programs in the Public Tier access data through public REST-based APIs. This capability means that researchers do not have to rely on us, as the originating developers, to construct tools.

The OSP must provide researchers with maximum freedom in their choice of tools.[3] So this thrust makes several APIs available, including the _Tool API_. The OSP will expose data using a REST-based[4] interface, which is a standard HTTP transport, and therefore is ubiquitously supported by programming languages and network infrastructures.

The OSP must provide researchers with maximum access to data on the OSP. _Researchers can use the Tool API for direct access to data._ Under REST, each dataset has a URL. The URL structure is human readable, and is easy to use. Dimensional data can be cut into ranges. For example:

- https://[study handle]/datasets/datasetName/samples/sample_id
- https://.../matrix/2500..60000/17..8500/
- https://.../movies/LifeOfBrian/17m3s/

The Operations Tier (shown in the middle of Figure 5) is the heart of the OSP. Most of the modular components in this tier are provided by this thrust; namely: the Sampler, Secure Views Model, Manager, and Job Scheduler.

For big data stores, the OSP will provide data samples. Often big data cannot be fully copied - at least for initial stages of research – to a local machine and should thus be sampled. Where applicable, _the Sampler_ will provide samples of data stored on OSP. The resulting "summary" URL will return a sample of the data (e.g. by using a hardware-based random sampler, or a Gibbs sampler[5]). A data sample could have multiple URLs, if it can be accessed from different indices. HTTP (and thus, REST) allows software to specify requested formats, using the ACCEPT header. OSP will support, where applicable, CSV, TSV, XML and JSON. Secure access control mechanisms will be used to protect sensitive datasets. The ability to access a single sample from virtually any programming environment allows researchers to use the OSP freely for data access, even in a massively parallel manner, such as the popular Map-Reduce method.[6]

The OSP must make sure data and tools are only available to those researchers authorized for access. While many of the data and tools on the OSP will be openly and freely available, controlled access is necessary to support private workspaces, privacy-sensitive data and other uses of the OSP. The _Secure Views Model_ is responsible for enforcing access permissions. _The Secure Views Model will be an innovation over state of the art permission systems._ Unlike historical file systems, where permissions are at the file and directory level, permissions on the platform are on views of data, which may be enforced on values appearing in a data stream, on a table or in a file, or may be an entire table or file or groups of tables or files.

---

[3] The OSP will support common tools, such as Hadoop, Zelig/R, Java, Python and MatLab.

[4] The OSP will initially implement the HTTP REST services using PlayFramework 2.0, a novel web framework in use today by many high-traffic, internet size websites such as LinkedIn and theGuradian.co.uk. Among other features, it is geared towards developing REST-based APIs, and allows HTTP streaming, a feature we plan to relay on for delivering large results. Standard caches, load balancers and proxy servers will be used as well.

[5] http://www.jstor.org/discover/10.2307/2685208 Gibbs Sampling is a data mining technique that has become quite popular these days and is being used quite extensively in Health Sciences and Financial Services.

[6] To further speed access, datasets can be served using a content delivery network (CDN). While commercial CDNs are available, it may be possible to build a CDN using existing campuses throughout the US and abroad.

The OSP must provide the multitude of features popularized by Dataverse and must additionally include provenance tracking for data. The *Manager* is the engine of the OSP, responsible for coordinating the storing, archiving, publishing and accessing of data and for tracking provenance. It contains the storage URL of each dataset and relevant API documentation and allows researchers to: discover existing datasets through manually cataloged fields or automated extractions; reuse data through the creation of persistent URLs to the stored data and automated data management; validate prior results using stored provenance information, as well as code and process documentation; and provide citation and branding features.

The OSP must support a plethora of data storage systems, local, remote, and standalone. The Data Tier (shown on the bottom) refers to a distributed, amorphous configuration of data storage systems, including combinations of cloud and local storage capabilities. Data storage is not done using a single system, but rather by various systems, allowing the data author to choose from among choices recommended by the Manager, the storage system best fitting the structure of the data and expected queries. Building on diversified data storage allows the OSP to add new and different storage types and to take advantage of standalone repositories and infrastructures (such as proxy servers and content delivery networks) in order to speed up data access.

The OSP must provide robust efficient data access. To support read-intensive computations, OSP will support batch processing of jobs *using in-database analytics* (*"in-DB Jobs"*). This concept leverages database kernels to perform various types of analytical computation, thereby placing computation near the data to reduce network latency. The end result is that large volumes of data can be analyzed seamlessly using MapReduce without having to ever extract data from the database. Researchers submit binaries of supported languages/frameworks, and get the results when operation completes. The *Job Scheduler* in the Operations Tier coordinates the execution of "in-DB Jobs". This thrust will provide in-DB data mining methods such as ANOVA, Linear and Logistic Regression, and Clustering for its local data storage systems.

*In summary, this development thrust will construct the basic operations of the OSP.* It heavily relies on existing systems and trusted standards, providing transparency while keeping costs low and allowing maximum flexibility to researchers.

## 2. Privacy-Sensitive Data

The objective of this development thrust is for researchers who work with privacy-sensitive data to enjoy all the data and tools on the Open Science Platform with certified privacy protection. Achieving this objective will be a significant scientific contribution.

The state of the art for researchers who work with sensitive data is to sequester the data on computers that are not on any network. Sometimes the sequestered machine may further reside in an isolated facility. As a consequence, it can be difficult to link the data with other information unless the researcher can obtain an independent copy and store it on the sequestered machine. Storage and computational options are limited to those available on the sequestered machine. This thrust will ensure that researchers who work with many forms of privacy-sensitive data can enjoy maximum benefit from resources on the OSP by safely using the data on the OSP.

There is no one privacy standard, so the OSP must support a full range of options from data having no privacy risk to data requiring maximum protection. We offer the following approach. Assign a privacy policy to each piece of data in the OSP, even if the data has no restrictions. Then, enforce policies throughout the system. While the space of possible policies is very large, complexity reduces because actual data handling requirements associated with policies collapse to few possibilities.

*We introduce a new 6-level privacy scale for the Open Science Platform to use to specify privacy and security policies for data.* This scale originates from the Harvard Research Data Security Policy, which defines a 5-level categorization schedule for research information and defines the minimum protections required for each level.[7]

Level 1 data is non-confidential information that can be stored and shared freely. Level 2 data may have individually identifiable information but disclosure would not cause material harm. Level 3 data may have individually identifiable information that if disclosed could be expected to damage a person's reputation or cause embarrassment.

Level 4 data may have individually identifiable information that includes Social Security numbers, financial information, medical records, and other individually identifiable information. Regulations often apply to these data, such as the Health Information Portability and Accountability Act (HIPAA) and the Family Educational Rights and Privacy Act (FERPA), which may dictate handling and sharing.

| Risk Level | Storage & Transit | Access |
|---|---|---|
| 1 No risk | Clear | Open |
| 2 Foresaken | Clear | Email/OAuth verified registration |
| 3 Shamed | Encrypted | Password, Registered, Approval, Data Use Agreement |
| 4 Civil Penalties | Encrypted | Password, Registered, Approval, Signed Data Use Agreement |
| 5 Criminal Penalties | Encrypted | 2-factor authentication, Registered, Appoval, Signed Data Use Agreement |
| 6 Max Control | Double Encryption | Password, Registered, Approval, Data Use Agreement |

**Figure 6. Six privacy levels and associated security features.**

Level 5 may have individually identifiable information that could cause significant harm to an individual if exposed, including serious risk of criminal liability, psychological harm, loss of insurability employability or significant social harm.

---

[7] http://www.security.harvard.edu/research-data-security-policy

Level 6 data requires an additional commitment that administrative and technical staff cannot inspect its contents. We expect to enforce this using double encryption with the owner's password being one key and the OSP having the second key, though working with opaque binaries will be a challenge that may limit some uses.

Figure 6 summarizes our initial privacy levels for the OSP with associated security features for each level. Examples of access requirements include: users having passwords; registrations being verified by an external email account (or OAuth); and, access requiring 2-factor authentication where a one-time password is sent to a device like a registered mobile phone. Approvals may be from an IRB, the data's curator, or those controlling the encryption key. One of many possible data use agreements may be required.

Certifying that the system handles data properly with respect to stated privacy standards relies in part on formal properties enforced by data handling. To certify that those standards harmonize with regulations, we will leverage lessons learned from Sweeney's earlier work [60-65]. Her Privacert model is the leading approach for certifying data as being sufficiently de-identified under HIPAA, for example.

*In summary, this development thrust will enable researchers to fully use the Open Science Platform with privacy-sensitive data.*

⌘ Related Activities Underway

Dataverse currently stores Level 1 and 2 data only. Sweeney, Crosas and others are already constructing a wizard that researchers will use to answer a series of questions to determine the proper classification of data being deposited. Figure 6 summarizes the current research state of associations of privacy levels to security features.

## 3. Sustainability

The focus of this thrust is to develop methods for making the Open Science Platform financially sustainable. The goal is to minimize the dependency of the OSP on research sponsorship without sacrificing open access to researchers. We will conduct a series of experiments to determine the financial viability of the OSP.

One of the experiments involves MyDataCan, where people assemble copies of their personal information and run apps on mobile or web devices to use their data to improve their lives. This approach is similar to Apple's iTunes. Third-parties can write apps for MyDataCan and charge a fee. The expectation is that most apps will be at no charge, but for those charging a fee, if a tax is kept by the OSP, would it be sufficient to cover the operating expenses of the OSP?

Another experiment involves charging access for curated high-profile research data, which itself has a recurring expense. An example is the public health data described

**Data Partner: MyDataCan**

MyDataCan.org is a pilot project at Harvard whose aim is to be a living lab for personal data sharing. Goals as a data partner are to incorporate MyDataCan into the Open Science Platform so that members of the public will have an opportunity to directly contribute their data to science and help render the OSP financially sustainable.

Using the PPR Trust Framework [74], which prescribes 65 rules of conduct when handling consumer information, the OSP will conduct an experiment with MyDataCan on the OSP to see if at least 100K Americans will store data in the repository and at least 50K of them will share some data with research. Additional experiments will determine whether funds acquired from taxing apps for MyDataCan can help financially sustain the Open Science Platform.

The goal of MyDataCan.org is to help a person gather a copy of the data he leaves behind, place it securely under his control, and then provide him with many possible ways to use and share his information. It uses the notion of a personal "can" of data over which he has *personal access control*. With his permission, data can be combined across "cans" " to enable research and to share data for many worthy purposes, inspiring puns (e.g. "MyDataCan Save Lives"). Not all decision making is left to the person. The design and governance of MyDataCan has rules that govern behavior out of paternalistic concerns to guarantee some protections. If successful, MyDataCan could be a tremendous resource for society, providing unparalleled access to personal data, and while doing so, introducing new ways of thinking about privacy in a data-rich networked world.

A participant gets value from MyDataCan by being able to assemble his personal information across disparate data sources, annotate his data, control distribution of his information, and receive personalized services made possible by apps that run on mobile and web devices that access personal information stored on MyDataCan/OSP. Simultaneously, MyDataCan provides a platform for researchers to solicit testbed subjects to participate in many possible research activities (e.g., clinical trials, public health surveys, health outcome studies, etc.) though no research participation is required.

The system has been piloted locally with a small number of participants storing GPS, video and medical data and using apps to display co-locations of friends and family, do medication reminders, and provide an emergency peace of mind application.

Data sharing occurs through an applications interface as the owner of the data permits. Many apps are likely to be programs that use the person's information to help him, such as a prescription refill reminder (Rx Refill) or an emergency room app that reports allergies and medications and an emergency contact if the person finds himself in an emergency room (ER Peace of Mind[8]).

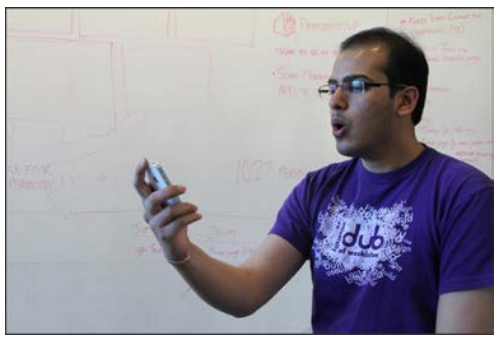

---

earlier. An annual fee of $500K is paid to the original source of the data, Medicare. Can that cost be recovered from researchers who access a derivation that has been infused with environmental data?

*In summary, this development thrust will help assure the long-term viability of the OSP. The target is to fund the Open Science Platform in the 5th year.*

## 4. Big Data Storage Systems

This development thrust focuses on data storage in the Open Science Platform, especially big data storage schemes. The goals for this thrust are to provide a broad array of storage systems on the OSP and to provide tools to help the OSP and researchers identify which storage system best fits the data, depending in part on size, but more often on function or intended query. Having an automated or semi-automated means of determining which storage system is best for a researcher's big data will be a scientific contribution. The state of the art is to make storage decisions primarily by type of data, not by kind of analysis. The wrong decision can dramatically slow performance.

Commercial big data systems such as Netflix, Amazon, Facebook and Flickr, have shared parts of their design and technologies with the public. We draw, in part, on what they have published and public availability of instances of their approaches to make these storage systems available to researchers.

Wide Column Store: fits data whose samples can be modeled using key-value pairs (e.g. {*name*: "Dan", *age*:12}), or that has a very intensive insertion rate. OSP will use Apache Cassandra[9].

Document Store: fits data whose samples need to be modeled using nested hash tables (e.g. {*name*: {given: "Dan", *surname*: "Smith"} *age*:12}). OSP will use MongoDB[10] to store datasets of this type.Graph Store: a good fit for storing data modeled as a graph, with annotated edges and vertices. OSP will use Neo4j[11] (currently used by the Dataverse Network) and will experiment with graph processing engines: Pregel or GraphChi and PowerGraph.

The Open Science Platform must have privacy-sensitive versions of big data storage systems to accommodate big data that are to be protected. Figure 7 enumerates storage options, as they relate to data exemplars and privacy levels (Figure 6). We will implement the full range of privacy levels on each big data storage system. Achieving this objective will be a significant scientific contribution. A challenge is using opaque binaries as key values, because encryption alone is not sufficient for privacy. For example, assume a dataset has two values for a field, A or B with distributions 25% and 75% then those distributions remain even if the values are encrypted.

*In summary, this development thrust will provide a range of big data storage systems for the Open Science Platform.*

---

[9] http://cassandra.apache.org
[10] http://www.mongodb.org
[11] http://www.neo4j.org

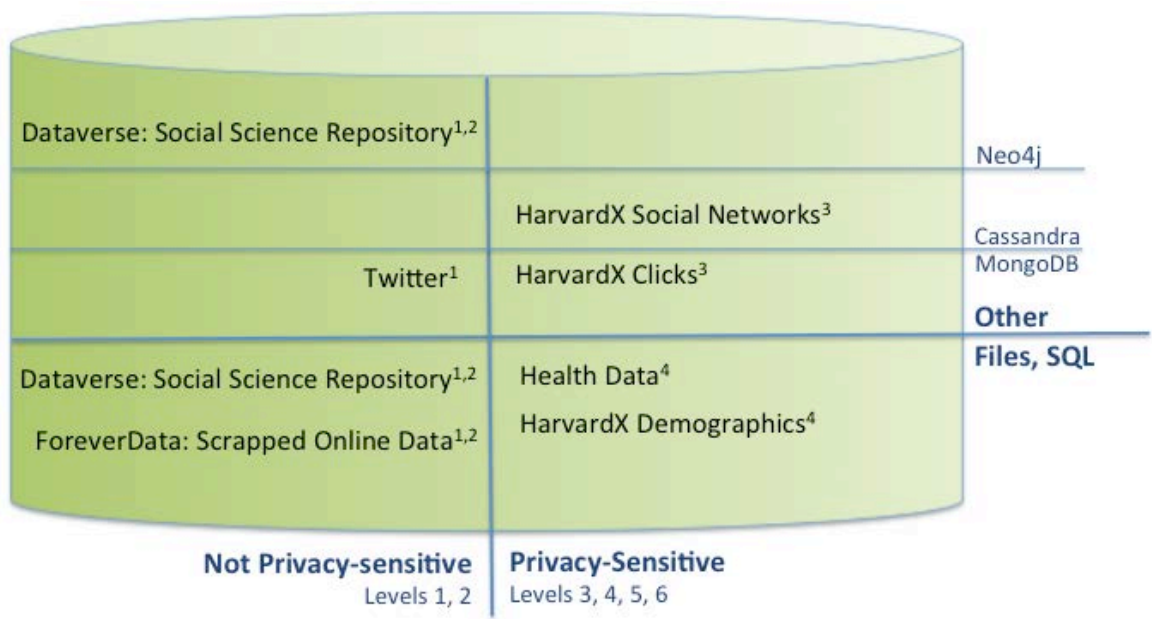

**Figure 7. Types of storage by data use (vertical axis) and privacy sensitivity (horizontal axis). Superscripts are levels.**

## 5. Data Provenance

This development thrust focuses on the provenance (or lineage) of data in the Open Science Platform. The goals for this thrust are to make provenance first-class meta-data in the OSP and to expose such data through a rich query interface.

The OSP will provide a provenance store that integrates provenance produced externally with provenance produced on the platform, exposing the combined data set through a rich query interface. Achieving this objective will be a significant scientific contribution. For example, researchers will be able to easily identify all the data sets that derive from their own contribution; they can identify the data sources on which their analyses depend; they can determine precisely what versions of what tools were used to produce a research result; and they can search for particular analytical processes.

Data in the provenance store will come from five different sources: original meta-data entered during data ingest, pre-existing provenance for data ingested from systems that already maintain provenance, coarse grain records of program execution, derived provenance that can be obtained by transforming existing tool output (e.g., R history logs), and data provided by provenance-aware tools.

When ingesting data, the OSP will require that users provide minimal metadata; this will serve as the most primitive form of provenance we store. In some cases, data will arrive

**Data Partner: Public Health**

<u>Public Health Researchers</u> need massive amounts of data to detect health effects of environmental agents and sophisticated statistical approaches to extract reliable evidence from observational data.

Dr. Dominici develops statistical methods for integrating and analyzing large and heterogeneous datasets to routinely evaluate the health consequences of environmental interventions [33-44]. A key accomplishment of Dr. Dominici is the National Morbidity and Mortality Air Pollution Study (NMMAPS) [45-56], a highly influential and frequently cited nationwide time series study of the effects on mortality and hospitalization of air pollution. To promote a free exchange of ideas and to influence environmental policy, <u>she made the entire NMMAPS database and accompanying open-source software freely available</u>[12] <u>However, a key obstacle to data sharing is privacy and concerns eventually led to these data being removed from public availability in 2011.</u>

Now her team has <u>established an unprecedented environmental health dataset</u>, called the Medicare Air Pollution Study (MCAPS), the largest nationwide retrospective cohort study and multi-site time series study of fine particles ($PM_{2.5}$) and hospital admissions for cardiovascular and respiratory diseases from Part A Medicare data [42,43,45,51,52,54]. Findings from NMMAPS and MCAPS have had a major impact on air quality policy regulation for particulate matter and ozone [35,45-59]. So, her team seeks to work with the data on the OSP and <u>release sufficiently de-identified versions of the data on the OSP</u>.

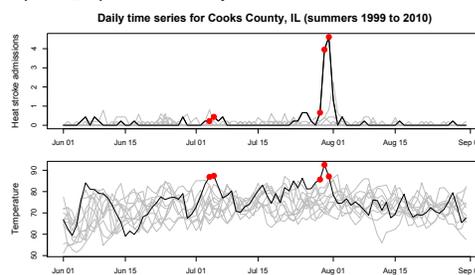

from a system that already captures provenance. In these cases, we will accept that provenance in a standard format, e.g., W3C PROV-XML, and incorporate it into the store. Whenever researchers produce data sets on the OSP, the identity of the input data sets and the identity and versions of the tools used to produce the derived data are added to the provenance store. In cases where the tools used provide sufficient information, such as R history logs, transformed versions of that data will be added to the provenance store. Finally, the OSP will make available easy-to-use programmatic APIs [20] that tool developers can choose to use to provide enhanced provenance that provides fine-grain detail about the data derivation process.

The provenance store on the OSP will provide several capabilities not found elsewhere. As outlined above, it provides a new way for users to track the use of their data products and seek information about other data products or search for the existence of particular experiments or data product. In addition, full provenance records provide a means to identify data collections that may need to be updated due to a new release of software or to modification of a source data set. Provenance of temporary collections can be useful in constructing persistent citations to temporary data sets. Critical challenges are the scale of the data and the level of details captured.

---

We leverage prior work by Seltzer in designing and implementing different provenance collection, maintenance, and query systems.

*In summary, this development thrust will result in the Open Science Platform having both the largest digital provenance collection of which we are aware and the most comprehensive, providing a long-term, detailed record of the scientific process.*

| ForeverData – Uploads of Harvested Online Data |
| --- |
| Foreverdata.org is a pilot project at Harvard.  The site harvests online data in set time intervals using scripts.  Examples include online data from twitter, voter lists, mug shots, and online ads. These data feeds will automatically be uploaded to the Open Science Platform using the Batch API. |

## 6. Data Citation for Big Data and Data Streams

This development thrust focuses on data citation from big data and data streams in the OSP.  As described earlier, the Dataverse Network has already accomplished reliable and persistent citation of fixed-size data sets, primarily by making copies as needed for preserving and curating the data.  But big data and big data streams do not lend themselves to making copies per citation. Instead, what is needed is a new way of making sure the same data values are captured and preserved so that results can be replicated. The goal for this thrust is to tackle this problem.  Achieving data citation for subsets of big data and data streams will be a significant scientific contribution.

The state of the art is the Dataverse Network, which automatically generates a formal data citation for each data study in a Dataverse based on the citation standard published by Altman and King [18]. The standard offers proper recognition to authors and distributors as well as permanent identification through use of global, persistent identifiers (the Dataverse uses handles through Handle.net to register permanent url [83], and will shortly also support DOIs registered through DataCite [84]). The standard includes a universal numerical fingerprint for quantitative data sets. Four features make the UNF especially useful: The UNF algorithm's cryptographic technology ensures that the alphanumeric identifier will change when any portion of the data set changes. Not only does this assure future researchers that they can use the same data set referenced in a years-old journal article, it enables the data set's owner to track each iteration of the owner's research. When an original data set is updated or incorporated into a new, related data set, the algorithm generates a unique UNF each time. The UNF is determined by the content of the data, not the format in which it is stored. Knowing only the UNF, journal editors can be confident that they are referencing a specific data set that never can be changed, even if they do not have permission to see the data. In a sense, the UNF is the ultimate summary statistic. The UNF's noninvertible, cryptographic properties guarantee that acquiring the UNF of a data set conveys no information about the content of the data. Authors can take advantage of this property to distribute the full citation of a data set - including the UNF - even if the data is proprietary or highly confidential, all without the risk of disclosure [85].

**Data Partner: Connectome**

Connectomics is the science of producing "wiring diagrams" of brains. These detailed maps of neuronal processes and synaptic connections show neural circuits at the nanometer scale. This research promises important insights into healthy cognition, as well as processes such as autism, epilepsy, and degenerative brain disease. Current research faces many challenges in storing, processing, sharing, and archiving large-scale data. Sizes of the datasets are in the petabyte range. Current experiments produce data at a rate of about 1/2 terabyte per day.

Professor Pfister, Dr. Knowles-Barley, and Dr. Jones want to use crowd sourced correction and annotation of automatically labeled connectome datasets on the OSP, which will require random access to small subsets of the data stored on XSEDE (Gordon ION Cluster). They use automatic analysis of serial section electron microscopy [21-31], but manual labor is required to improve accuracy in 3D neural reconstructions, specifically to proofread and make corrections, with results feedback to the processing algorithm [32] running on XSEDE so as to reduce errors even further.

The Connectome team will setup the tools and connections needed through the OSP and evaluate performance.

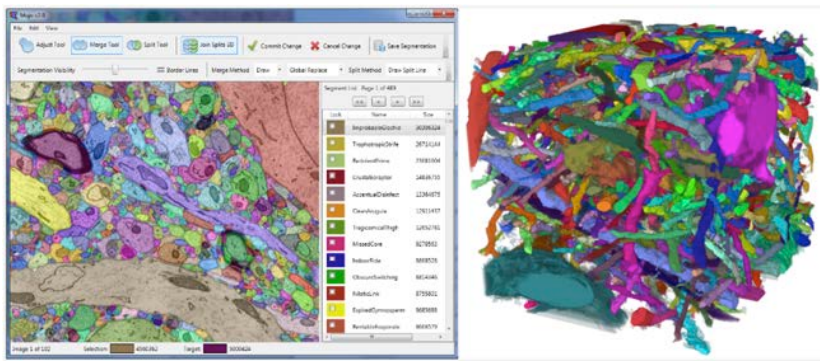

A full data citation for a quantitative data set in Dataverse is as following:

Nielsen, Richard, 2010, "Replication data for: Foreign Aid Shocks as a Cause of Violent Armed Conflict", http://hdl.handle.net/1902.1/20243 UNF:5:req1o9CFaq3IxaFPXvEJyQ== V4 [Version]

An initial insight is to just provide citations for samples from big data and from data streams following the same guidelines described above, but to expand the citation coverage to provide a data citation for the entire data set, as well as to generate on the fly a data citation for a sample of the data set or stream. The persistent url associated with the sample will resolve to a citation RESTful API that will extract the reproducible sample from the entire stream. The landing page will include metadata describing the sample, as well as description of the entire data set.

The new data citation tool will also calculate a fingerprint on the values of the sample at the time that the data citation gets generated.

For images (e.g., TIFF, JPEG2000) and graphs (e.g, GraphML), the new data citation tool will calculate a cryptographic fingerprint following similar rules applied to the UNF for quantitative data sets, but adapted to work for standardized images and graph files.

**Data Partner: Astronomy**

Astronomers use Pan-STARRS, a synoptic sky survey. Its goal is to image large swaths of the sky at a high time-cadence, so as to look for time-varying astrophysical sources.  At present, the Pan-STARRS static data are arranged in a custom, highly parallel (map-reduce) python-based database called the Large Survey Database (LSD). It scales well to tables of $10^{11}$ object detections, and scales well on a Beowulf cluster.  The parallelization makes it fast; as a result, an astronomer's ability to work the Pan-STARRS data. At present, the discovery of transient sources in the Pan-STARRS data necessitates several hours per day on the school's computer cluster, while the identification of genuine sources (vs. image artifacts) and their classification is left to manual techniques.  This effort is not sustainable with the next generation of surveys, which will produce 10-100 times more data.

Professor Goodman, Dr. Finkbeiner and Dr. Berger anticipate that machine-learning techniques can help solve identification and classification problems in the existing Pan-STARRS database. If so, astronomers could reject the large fraction of false positive signals, and robustly classify the genuine transient and variable sources into distinct categories.  Such an effort will need to be carried out in real time as data flow in on a nightly basis. Unfortunately, much of the expertise needed in machine-learning algorithms resides outside of the astronomy community, so this data partner wants to use tools (e.g., MapD) on the Open Science Platform.

This data partner will also add a tool to the OSP. Glue[13] is a python-based visualization tool that allows its users to visualize and inter-compare several data diversely-formatted sets at once, without creating a joined data file.  For an astronomical example, imagine that: a.)  a virtual observatory tool offers an infrared map of a nebula; b.) a researcher using theastrodata.org has posted a catalog of young stars near that nebula; and c.) another researcher used a VO-enabled National Radio Astronomy archive to post new maps of gaseous jets from some of the young stars.

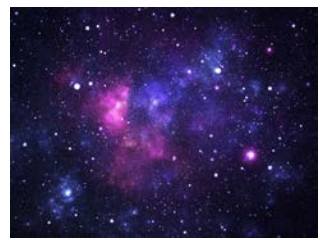

This approach can work as long as the researcher acquired a complete copy of the data. But more often, researchers may use tools like MapD with a sample is grabbed in real-time and not retained (e.g. the HarvardX data).  In fact, it may be impractical to retain a copy of all samples of billions of values a researcher explores. In these cases, a new approach is needed. Another challenge is finding a canonical form for fingerprinting images independent of storage format (e.g., the Connectome data).

We leverage prior work by Dr. Crosas and Professor King in the design and implementation of citation systems, especially that used in the Dataverse Network described earlier.

*In summary, this development thrust will allow researchers to cite and replicate data taken from big data and data streams and other data types such as images and graphs on the Open Science Platform.*

## 7. Data Exploration using MapD

---

[13] http://glueviz.org

This development thrust will develop an ultrafast analytical tool for exploring data on the Open Science Platform. The goal of this thrust is to develop an analytical tool that harnesses the massively parallel architectures of modern graphics cards to provide ultra-fast visualization and analytic solutions for datasets of all sizes, local or remote. Achieving this goal will dramatically impact the way researchers explore data. This thrust uses Graphics Processing Units (GPUs) to perform analytics on large volumes of data with unbelievable speed.

Analyzing large volumes of data, big data or data streams is a challenge. One cannot use conventional methods of extracting data from the databases and thereafter analyzing it using analytical software like SAS, R, MatLab, etc. The challenges with conventional approaches are that it is time consuming to move large volumes of data and secondly, if one can even move large volumes, traditional analytical software cannot handle large volumes on account of memory constraints.

Todd Mostak and Benjamin Lewis have already developed a prototype at Harvard called MapD (Massively Parallel Database) that takes advantage of the immense computational power and memory bandwidth available in commodity-level, of-the-shelf graphics processing units, originally designed to accelerate the drawing of 3D graphics to a computer screen, to form the backbone of a vertically-integrated data processing and visualization engine that marries the data processing and querying features of a traditional database system with advanced analytic, visualization, and machine learning features.[14] MapD is designed to run on hardware configurations ranging from GPU-equipped laptops to dedicated 16-card GPU clusters, achieving many orders of magnitude speedups for common workloads.

Even on a small GPU server, MapD can query and visualize billions of streaming data points in real time (i.e., with latencies measured in milliseconds). As the name suggests, MapD has been designed to allow for on-the-fly geospatial exploration of large datasets, including the generation of point maps, heat maps, and choropleths that can be consumed by any web client. In addition, modules are currently being developed that will allow MapD to accelerate standard OLAP datacube operations such as those provided by systems like Teradata and Vertica, making them real time as well.

Figure 8 shows a real-world example of a MapD-generated heatmap, point map, and time graph against hundreds of millions of tweets. Note that pointmaps are also generated using a binning and convolution filter. The time to generate this map using a small MapD instance was less than 50 milliseconds, which included evaluating the query (a text search for "flu" across hundreds of millions of records, rendering these results to a texture, and finally, in the case of the heatmap, spatially aggregating the resultant texture using a Gaussian convolution filter).

---

[14] MapD powers Worldmap, available at http://worldmap.harvard.edu/.

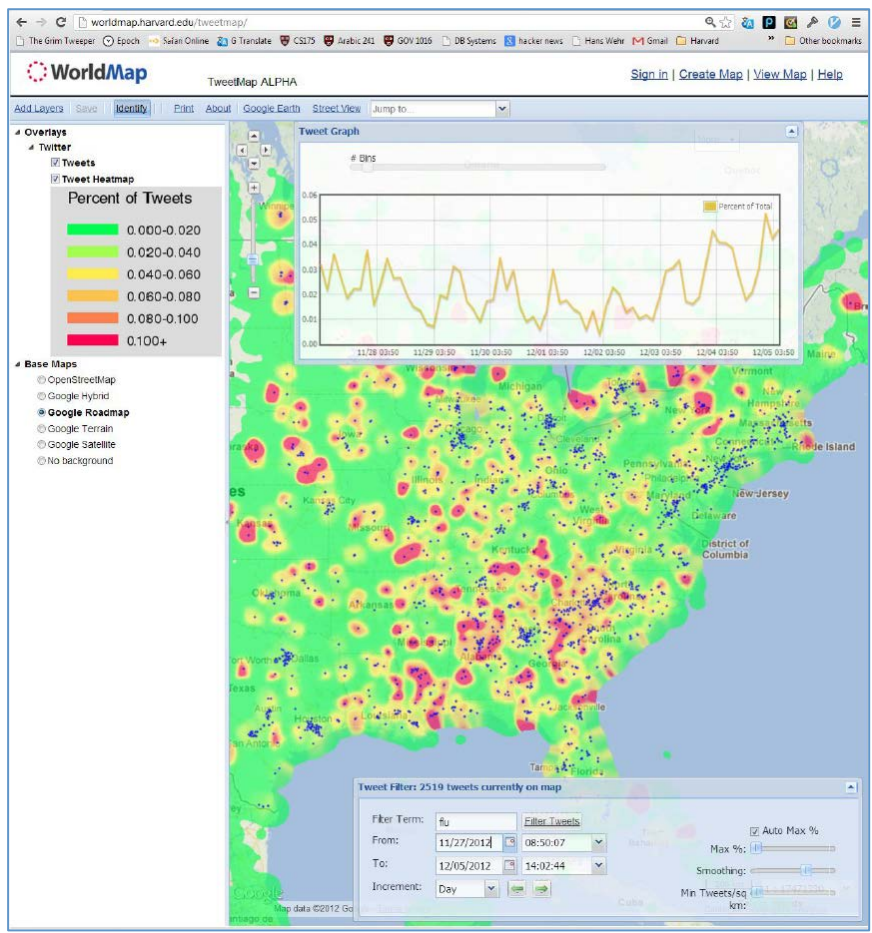

**Figure 8. Example of an on-the-fly visualization done by MapD using hundreds of millions of tweets.**

For the Open Science Platform, this thrust will develop fast, interactive, visualization interfaces to allow users to slice data across different dimensions simultaneously to facilitate querying, analyzing, and visualizing of big data.  Response times will be fast enough to support on-the-fly temporal and spatial fly-throughs of the data using commodity hardware.

All software written for MapD and for OSP will be open-source, with communication between browser and server handled by HTTP.

Because the proposed visualization software/hardware platform is inexpensive it can be used to support flexible and evolving distributed or centralized (or hybrid) architectures, and because it can scale, multiple MapD instances can be chained together to support larger datasets or more intensive processing tasks or greater user traffic.  Depending on the type and volume of processing required, chaining is possible over low, medium, or high bandwidth connections.

Any dataset can be visualized using MapD.   Given the memory capacities of current GPU cards and current standard server configurations, a single server could provide visualization and analytics against several billion features.

An instance of MapD resides on the platform supporting visualizations against the large number of relatively small datasets of the entire Dataverse Network, while the big data participants in this proposal (astronomy, neuroscience, health, HarvardX, Twitter) would each have a dedicated MapD instance to support visualizations for their datasets. Additional instances of MapD would be available for scheduling through XSEDE or the Job Scheduler on the OSP. Communication between each remote MapD instance would be via a RESTful HTTP, as described by the Batch and Data APIs on the OSP (discussed earlier in the Core Platform section). For datasets to be ingested by MapD, they will need to be stored in a supported format.[15] For large datasets, metadata for the dataset might optionally be loaded into MapD as well, allowing, for example, for the efficient exploration of large binary "blob" datasets such as the mouse brain TIFFs currently stored in the neuroscience connecteme data.

This thrust will also develop a full-featured browser-based data dashboard tool for researchers to interface with MapD allowing for real-time querying, visualization, and analysis of diverse datasets on-the-fly, having data query, visualization, and analysis. In addition, this thrust will provide a version of WorldMap through the Tool API on the OSP. WorldMap allows researchers to explore, visualize, edit, collaborate with, and publish geospatial information on the OSP.

*In summary, this development thrust will allow researchers to do on-the-fly temporal and spatial fly-throughs of data on the Open Science Platform.*

## Conclusion

Direct impacts of this project are numerous. Researchers worldwide gain support throughout the research lifecycle and across scientific disciplines through the broad availability of research data and tools. Developers worldwide can share software tools. Other data repositories can expand their data sharing. Publishers gain a framework for citing data and reproducing results. Emerging researchers at the high school, undergraduate and graduate levels as well as the public and citizen scientists have participation opportunities in research. And individuals across the country can privately store and manage their own personal data, with an opportunity to consent to share some or all their data for scientific research.

More specifically, educational and outreach activities include:

- Targeted to the broader research, publishing and funding community: Two conferences (2nd year and 4th year), hosted at Harvard University and aimed to about 100 attendees each, on the topic of Open Data, Reproducible Research, and Open Data Exploration.

---

[15] CSV, Solr, PostgreSQL, MySQL, Json, GraphML, R, shape files, STATA, SPSS, TIFF, CSV, and R files and connectors for Solr, PostgreSQL, MySQL, and shapefiles are already supported in the MapD prototype.

- <u>Targeted to undergraduate and graduate students at Harvard University</u>: A series of courses through the Harvard College's Winter Session Program (which offers extracurricular winter break courses) to educate students on data sharing and contributing to the open research platform. All course materials will be made available online and open to students from other Universities.

- <u>Targeted to undergraduates worldwide</u>: The creation of an online HarvardX course ([https://www.edx.org/university_profile/HarvardX](https://www.edx.org/university_profile/HarvardX)) to disseminate how to use the open research platform for conducting reproducible research with confidential data. Course topics will include: 1) what is reproducible research, 2) technology tools for enhancing reproducible research, 3) barriers to reproducible research, 4) case studies of reproducible research (for example, air pollution and health). HarvardX is part of the EdX enterprise founded by Harvard University and the Massachusetts Institute of Technology. About 50K students worldwide might participate in a HarvardX course simultaneously.

- <u>Targeted to High School students</u>: Partnerships to provide high school internships to participate in the project.

- <u>Targeted to a wider audience</u>: A series of YouTube videos with interviews with the collaborators and introductions to open science, reproducible research and the scientific data accessible through the open research platform.